# Structural attributes and photo-dynamics of visible spectrum quantum emitters in hexagonal boron nitride

*Nathan Chejanovsky [1,2], Mohammad Rezai [1], Federico Paolucci [2], Youngwook Kim [2], Torsten Rendler [1], Wafa Rouabeh [2], Felipe Fávaro de Oliveira [1], Patrick Herlinger [2], Andrej Denisenko [1], Sen Yang [1], Ilja Gerhardt [1,2], Amit Finkler\* [1], Jurgen H. Smet [2] and Jörg Wrachtrup [1,2]*

[1] 3. Physikalisches Institut, Universität Stuttgart, Pfaffenwaldring 57, 70569 Stuttgart, Germany

[2] Max Planck Institute for Solid State Research, Heisenbergstr. 1, 70569 Stuttgart, Germany

**Abstract**

Newly discovered van der Waals materials like $MoS_2$, $WSe_2$, hexagonal boron nitride (h-BN) and recently $C_2N$ have sparked intensive research to unveil the quantum behavior associated with their 2D structure. Of great interest are 2D materials that host single quantum emitters. h-BN, with a band-gap of 5.95 eV, has been shown to host single quantum emitters which are stable at room temperature in the UV and visible spectral range. In this paper we investigate correlations between h-BN structural features and emitter location from bulk down to the mono-layer at room temperature. We demonstrate that chemical etching and ion irradiation can generate emitters in h-BN. We analyze the emitters' spectral features and show that they are dominated by the interaction of their electronic transition with a single Raman active mode of h-BN. Photo-dynamics analysis reveals diverse rates between the electronic states of the emitter. The emitters show excellent photo stability even under ambient conditions and in monolayers. Comparing the excitation polarization between different emitters unveils a connection between defect orientation and the h-BN hexagonal structure. The sharp spectral features, color diversity, room-temperature stability, long lived meta-stable states, ease of fabrication, proximity of the emitters to the environment, outstanding chemical stability and biocompatibility of h-BN provide a completely new class of systems that can be used for sensing and quantum photonics applications.

**Main Text**

Hexagonal boron nitride (h-BN) has a honeycomb lattice structure, with space group $D_{4h}^6(P6_3/mmc)$.[1,2] B-N atoms have a nearest neighbor distance of 1.4 Å.[3] The 2D layers consist of $sp^2$ covalent bonds with an alternating stacking pattern of boron and nitrogen atoms.[4] As opposed to graphene, all nuclei in h-BN have non-zero spin. The Raman Stokes shift of h-BN is in the range between 1363 $cm^{-1}$ and 1370 $cm^{-1}$, depending on the flake thickness.[1,5]

Recently, single quantum emitters (SQEs) residing on the edges of WSe$_2$ flakes have been discovered for temperatures below 15 K.[6,7] However, the bandgap of WSe$_2$ is relatively small (1.35 eV)[8] compared to that of h-BN (5.95 eV),[3,9] making it more difficult to probe with intra-bandgap excitation. h-BN is therefore a natural candidate for optical defects research like 3D materials that host optical defects, such as diamond which hosts e.g. the nitrogen-vacancy (NV) center. [10] A major difference, however, is that because of the 2D nature of h-BN, a defect would be closer to the external environment. This results in a heterogeneous environment in which a defect in h-BN resides. As a consequence, we expect the photoluminescence (PL) spectrum, lifetimes and other related phenomena to reflect this heterogeneity.

h-BN has been widely studied for its emission in the UV spectral range [11] and was recently proven to have an indirect bandgap.[9] Exciton UV emission in the ~220 nm spectral range has been attributed to grain boundaries and dislocations [12] and, using cathodoluminescence, for stacking faults localized at fold crossings of the flakes, resulting in local symmetry changes.[13] For the ~300 nm range, nitrogen vacancies and carbon impurities have been ascribed as the source of emission.[14,15]

Recently, it was demonstrated that h-BN also hosts SQEs in the UV [16] as well as the visible range [17,18,19,20] (623 nm).[21] The defect thought responsible for quantum emission at 623 nm was an anti-site complex, i.e. a nitrogen occupies the boron site with a missing atom at the nitrogen site, $N_B V_N$.[21] The high density of emitters found,[21] in account with the fact that due to the 2D nature of h-BN the emitters are exposed to the environment, suggests that sample preparation should have a significant influence on its behavior.

In this paper we show two possible methods that can be used in order to create QEs in h-BN utilizing chemical etching and ion irradiation. We probe the QEs using 532 nm green laser and 594 nm orange laser, which should assist in exciting inside the band-gap and thus avoiding excitonic emission. We analyze the spectral behavior of numerous cases of QE from different sources from bulk down to the mono-layer and show the similarity between them and an orientational correlation to the structure of h-BN. An analysis of photo-dynamics reveals a complex picture, which can hint to the presence of charge states. From these methods and previous research, we try to deduce the possible origin of the SQEs involved. Our results suggest that SQE positional and transition rate behavior have a connection to the reduced dimensionality of h-BN and our analysis points to a possible link between perimeters and defect formation.

We investigate h-BN from several sources: exfoliated flakes, purchased mono-layer CVD h-BN (Graphene supermarket) and in-house grown CVD h-BN, comparing structural, spectral and autocorrelation similarities between them. Measurements were performed on $SiO_2$ substrates. For the exfoliated h-BN we also tested a diamond substrate - to exclude substrate contributions to the SQEs. Some of our $SiO_2$ substrates were etched with patterned holes to suspended h-BN flakes in order to avoid interaction with the surface. A microscope image is given in Fig. 1b.

In addition to testing various sample sources we also performed two treatments on the exfoliated h-BN flakes to generate defects.

The first treatment was chemical: two etching methods were tested on exfoliated flakes, which thin down the flakes and proved important for generating single emitters. The first method makes use of peroxymonosulfuric acid ($H_2O_2$:$H_2SO_4$), which can be tuned by changing the ratio and etching time to effect a slow rate etching of h-BN.[22] This is similar to standard methods that are used for diamond cleaning for use of the $NV^-$ defect. The second chemical etching method is based on the first one but with an additional step of phosphoric ($H_3PO_4$) and sulfuric acids ($H_2SO_4$). This additional step can thin down the h-BN flake at a slow and controllable rate [23] (see SI for further details). The second treatment involved ion irradiation: we irradiate the flakes with He and N ions to generate lattice vacancies, similar to procedures known from 3D materials.

All measurements were performed with a room temperature confocal setup where photon emission was recorded with two avalanche photo diodes (APDs) in a Hanbury Brown & Twiss configuration. The majority of confocal scans were done with a 532 nm continuous wave (CW) laser with a 550 nm long pass (550LP) filter. We limited our measurements to the low power density regime of 150 µW/µm². In addition, we also recorded images with low laser power (nW) without any filters in order to gain insight to the correlation between emitter location and the flake structure. In this very low laser power regime, emitters cannot be excited efficiently (as will be shown below), which allows us to see the structural features of the flake on the area in which they are embedded when scanning a confocal image.

We observed that prolonged (4 hours or more) exposure to the 532 nm excitation laser (in the high power range ~ mW) can cause blinking emitters to restore to a stable emission state (not shown). However, this method was inefficient and the mechanism behind it is unclear, perhaps due to local heating that can have a chemical bonding effect. Furthermore, annealing proved to be a crucial step for emitter stability in order to avoid working with blinking emitters and consequently for all flakes in this manuscript (treated and un-treated) we annealed the samples. The majority of samples were thermally annealed using the procedure described in Ref. 21 – under an argon environment at 850 °C.

In the following we first discuss our methods to generate quantum emitters in h-BN and then proceed to spectral and transition rate analysis of the QEs. We confirm the presence of h-BN using Raman spectroscopy (Fig. 1a) - suspended over a hole in the substrate (Fig. 1b). Fig. 1c /1d shows an unfiltered/filtered (respectively) confocal image of an area of the flake in Fig. 1b. suspended over a hole (hole marked as a purple circle in Fig. 1d). A suspended emitter can be seen in this area (marked in a red circle) in the depth (Z) and lateral (XY) scan. Thus, quantum emitters are not localized on the interfaces between the h-BN flakes and the substrate, but are an intrinsic property of the suspended flake. On the exfoliated flakes we see emitter densities of less than 10 emitters per *whole* flake (~30 µm lateral). We also notice a tendency of bright emitters to appear on the *perimeters* of the h-BN flakes: either on the boundary of a whole flake (Fig. 1e) or at the interface between flakes (Fig. 1f/1g, filtered/unfiltered, respectively) or clustering along lines inside the flake which show a constant angular multiplicity between them (Fig. 1h/1g), which, in this case, is $\alpha \sim n \cdot 30°$ (for details see SI). However, the confocal resolution is limited at best to 200 nm and hence we cannot conclusively say that the emitters are precisely positioned on the edges of these perimeters.

Subjecting the h-BN flakes to the chemical etching methods increased the emitter density dramatically compared to the exfoliated flakes. The number of emitters found per confocal area increased to 0.09 and 0.54 per $\mu m^2$ for both chemical methods using peroxymonosulfuric acid and the additional phosphoric acid ($H_3PO_4$) + sulfuric acid ($H_2SO_4$), respectively (see Table 1 for details). In Fig. 1j, a confocal scan of a chemical etching method using phosphoric acid is shown. The calculated density values are found in Table 1. More examples can be seen in the SI.

Deducing that an etching (atom removal) mechanism[22] is important for emitter creation, we also attempted ion irradiation on samples with nitrogen (N) and helium (He) on multilayer h-BN (See SI for details). We note that if ion irradiated samples were annealed at 850 °C in vacuum and not in argon, these emitters were less stable than those annealed in argon. Confocal scans of irradiation show higher density than the chemically treated samples. This high density was seen for both irradiations (see Table 1), emphasizing the role of inelastic collisions (vacancy creation) for emitter formation in h-BN. Although we are using N ions, $N_2$ ions have been shown to create interstitials in h-BN in addition to vacancies.[24] In addition, atomic substitutions cannot be discarded. We discuss below the possible effect of these. Due to the inert nature of the irradiated ions and low dimensionality of h-BN, it is less likely to implant them in the lattice and therefore we assert that the mechanism of emitter creation is not atom substitution.

Next, we investigated defects in mono-layers. To that end we compared the structural features and density of emitters in purchased mono-layer CVD h-BN to exfoliated flakes and to treated (chemical etched/ion irradiated) flakes. CVD h-BN is typically grown on a Cu foil substrate, where the first mono-layer growth is by the Frank van der Merwe model (i.e., layer-by layer) and afterwards changes to Stranski-Krastanov model (i.e., island on layer).[25] An SEM image of CVD mono-layer h-BN reveals non-uniform coverage already on a 20 μm scale (see SI). We use the same transfer method procedure described in Ref. 21 also on a 180 μm thick $SiO_2$ with patterned holes for commercially available graphene supermarket h-BN. Probing the sample with a 532 nm CW laser immediately reveals that emitter density (Fig. 1k) is larger than for chemically treated flakes, but lower than for high dose ion irradiated samples. Considering the inhomogeneous h-BN coverage on the Cu foil observed in the SEM image with our observations on the untreated and treated (chemical etched/ion irradiated) flakes, we correlate the high density of emitters with the large number of *perimeters* appearing in a CVD mono-layers due to the growth mechanism.

In Table 1 we summarize the h-BN types/methods for increasing emitter density.

**Table 1.** Summary of Types/Methods for emitter generation the dependency for each method and relative density of emitters per μm².

| Type/Method | | Exfoliation | Chemical etching | | CVD mono-layer | Ion implantation | | |
|---|---|---|---|---|---|---|---|---|
| | | | $H_2O_2:H_2SO_4$ | additional $H_2SO_4:H_3PO_4$ | | He | N | |
| | | | | | | Low Dose | High Dose | Low Dose |
| Density [$\mu m^{-2}$] | 532 nm Excitation | < ~10 emitters per flake (lateral ~ 30 μm) | 0.09 | 0.57 | 0.64 | 0.59 | 0.73 | 0.73 |
| Density [$\mu m^{-2}$] | 594 nm Excitation | ----- | ----- | ----- | 0.37 | ----- | ----- | ----- |

Densities are calculated for areas that showed the maximum number of emitters on a flake, a minimum distance of 0.5 $\mu m$ between emitters was taken into account.

We now discuss the spectral and autocorrelation properties of the various h-BN types and show that the spectral features resemble each other.

We start by analyzing a single quantum emitter (SQE) created using the chemical etching method with the additional $H_2SO_4$:$H_3PO_4$ treatment. Fig. 2a/2b shows a cross section of an area inside a flake where we compare between a filtered/unfiltered (respectively) image. The SQE is marked in the center of the red circle. We mark areas A and B on the image and measure the Raman signal. For A we get 1365 cm$^{-1}$ and for B we get 1369 cm$^{-1}$ after Lorentzian fitting (Fig. 2c), therefore indicating that this area is composed of thicker/thinner (respectively) h-BN.[5] We measured the emitter PL spectrum (Fig. 2d,right) and also show an autocorrelation measurement (inset), proving it is a single quantum emitter ($g^2(0) = 0.26 < 0.5$). At the end of our measurements we significantly increased the excitation power (~mW) to get a Raman signal and took higher resolution PL spectrum. This is shown in Fig. 2d – left. A close look reveals that the Raman signal is superimposed on our SQE spectral features. The inset displays the conversion to wavenumbers giving a value of 1367 cm$^{-1}$.

Fig. 2d - right shows a typical emission spectrum with an emission origin at 580 nm and two side peaks separated by a wavenumber differences of $1329 \pm 66$ cm$^{-1}$ and $1303 \pm 109$ cm$^{-1}$. These differences between the emission peaks in Fig. 2d are very close to the Raman shift of h-BN [1,5] (Fig. 2d upper scale). We therefore conclude that the central peak is the zero phonon line emission (580 nm), whereas the second one is the first Raman Stokes harmonic followed by the second one. Several PL measurements that were done on treated and un-treated h-BN samples have shown that this double peak structure appears on all emitters that exhibit quantum emission traits. However, the third peak is not always prominent enough to be seen. Nevertheless, the possibility of a second ZPL cannot be excluded and further measurements are needed for verification. Fig. 2e shows six PL spectra taken from QE on diamond and SiO$_2$ substrates, emphasizing that although emitters occur at different wavelengths, the double peak structure is recurring and is not substrate dependent. All of these six representative spectral features were seen both on diamond and SiO$_2$ and were not confined to the boundary case (Fig. 1e).

Polarized in-plane modes for h-BN at the Γ,M and K Brillouin points should have an energy of ~170 meV for both the transverse and longitudinal optical phonons (TO and LO modes), whereas polarization along the c axis is expected for energies below 100 meV.[2] The spectral difference in the emission origin can arise from strain (shown for SQEs in WSe$_2$ over a range of 170 meV [26]), different dielectric environments and most probably from at least two types of defects or different charge states for the same defect - we discuss later in detail the possible causes. We find a minimum full width half maximum (FWHM, denoted as ν) of 3.96 nm on the PL spectra for non-monolayer flakes.

We perform a similar analysis on 23 adjacent peaks from various sources of h-BN (CVD monolayer h-BN, treated (chemically/irradiated) and un-treated emitters). We plot the difference between adjacent peaks in wavenumbers as a function of standard deviation (Fig. 2f). The majority of wavenumbers are located close to the value of 1363 cm$^{-1}$ with an average value of 1337.6 cm$^{-1}$, very close to that of h-BN (1363 cm$^{-1}$ to 1370 cm$^{-1}$ depending on flake thickness).[5] We therefore show that the electron interacts with bulk or flake phonon modes rather than with localized vibrational modes as it is the case for defects which do comprise impurity atoms with mass significantly different from atoms of the host lattice. This spectral behavior was seen also for defects in h-BN that emit in the UV range.[11] The energy difference between the peaks fits the splitting between LO and TO optical phonons.[2] Finding similar modes for various defect types further demonstrates this behavior of interaction with delocalized phonons.

Ion irradiation produced similar spectral features like those in Fig. 2d, see SI for details.

We next examine the PL spectral features of several emitters in purchased mono-layer CVD h-BN. Fig. 3a shows spectral features of three emitters in a monolayer on SiO$_2$. As stated before, a similar Raman harmonic structure appears also in the monolayer. We see that these three emitters have ZPLs of 583 nm, 643 nm and 691 nm with energies in the range of less than 65 meV to those found in thicker flakes in Fig. 2e (the pairs of [569 nm, 594 nm], [622 nm, 650 nm], [683 nm, 697 nm]). For comparison, detuning up to 170 meV has been seen for the ZPL of emitters in WSe$_2$.[26] We find that sharp spectral features are also present in the monolayer regime, with an FWHM (ν) minimum of 9.94 nm. This is approximately a factor of 2 larger than the minimum found in thicker h-BN. Fig. 3b shows the spectral features of a suspended emitter with a ZPL of 588 nm, similar to that shown in Fig. 2d. We find that when probing the CVD monolayer with an orange CW laser (594 nm) defects in monolayers show photoluminescence for several tens of minutes (Fig. 3c). In contrast to measurements performed with green laser, where emitters show blinking behavior, with orange excitation this behavior is not seen on stable defects, although sometimes after 7 hours or more of excitation the emitters bleach, perhaps due to an ionization mechanism or oxygen quenching.[14,27,28] Near ZPL resonance, excitation stability was also seen for SQEs in WSe$_2$.[29] Fig. 3c demonstrates this stability for photon emission for a 20 µW 594 nm CW excitation for 45 minutes. In multilayer h-BN emitters could be probed more easily using the 532 nm CW laser and probing with 594 nm was not necessary. Fig. 3d shows PL spectra for four emitters on SiO$_2$ with ZPLs of 629 nm, 640 nm, 643 nm and 657 nm for excitation at 594 nm.

The different layer thicknesses from multilayers to a monolayer imply that the local environment the emitter is exposed to can be significantly different on the c axis (from mostly B-N atoms, to atoms of the substrate, in this case SiO$_2$). We find that for emitters on CVD monolayer the brightness is significantly different for SQEs which have almost identical ZPLs (in the range of 89 meV, Fig. 4a). This is surprising due to a more homogenous environment each emitter in the monolayer should have compared to emitters in multilayers. We therefore firstly analyze the photo-dynamics for multilayers to gain insight into the possible mechanisms involved. In Fig. 4 we analyze the power dependent photo-dynamics for three SQEs with ZPLs: 696 nm (4b), 580 nm (4c, 4d) and 639 nm (4e, 4g, 4h). The insets of Figs. 4b, 4d and 4h, depict our dynamics' scheme, for a two-level and a three-level system, using:

$g^{(2)}(\tau) = 1 - e^{-|\tau|\lambda_1}$ and $g^{(2)}(\tau) = 1 - (1+a)e^{-|\tau|\lambda_1} + ae^{-|\tau|\lambda_2}$, respectively, where $\lambda_1 = R_0^{21}(1 + \alpha P_{exc})$ and $\lambda_2 = R_0^{31} + \frac{R^{23}R^{12}}{\lambda_1}$. $\lambda_1$ is the excited state transition rate and $\lambda_2$ is the meta-stable state transition rate, $P_{exc}$ is the laser excitation power, $\alpha$ is a fitting parameter attributed to the power dependence of the excited state. We denote '1' as the ground state, '2' as the excited state and '3' as the meta-stable state where $R_0^{ij}$ is the transition between the $i \rightarrow j$ states for zero excitation power. We assume the inter-system crossing (ISC) $R^{23}$ is power independent. We note for emitters having $R_0^{21} > 500$ MHz the APD jitter can play a role in the measured lifetime.[30] For these measurements we fitted our data with a convoluted $g^{(2)}(\tau)$ function with a Gaussian function $J(\tau)$ representing our system response time, giving us the de-convoluted values. Power dependent autocorrelation measurements were done on each SQE. For SQEs that showed bunching behavior, in effect $g^2(|t| \sim > 5 \text{ ns}) > 1$ resembling long 'shoulder' like curves, similar to the $g^2(t)$ displayed in Fig. 2d (see SI for more details), analysis was done as a three level system and the decay $\lambda_2$ was extracted by fitting a decaying exponential function. Using the value of $\lambda_2$, the anti-bunching area was fitted to extract $\lambda_1$. $\lambda_1$ is plotted as a function of excitation power and fitted using the $\lambda_1$ function. We then extrapolate to zero power to extract $R_0^{21}$. The same procedure was done for $\lambda_2$ to extrapolate $R_0^{31}$ and $R^{23}$.

For emitter (696 nm ZPL) analysis was done as a two level system since no bunching behavior was seen on the autocorrelation curves. We get an excited state relaxation rate of $R_0^{21}$=1295±242 MHz.

For emitter (580 nm ZPL), shown in Fig. 2c, we get an excited state relaxation rate of $R_0^{21}$=303.98±10.30 MHz, a long-lived meta-stable state with $R_0^{31}$=162.12±92.45 Hz and an ISC rate of $R^{23}$=3.06±0.32 KHz.

For emitters in the monolayer (Fig. 4a) that showed bunching behavior (ZPLs 660 nm (black), 630 nm (red), 637 nm (blue) and 657 nm (green)) we note that the meta-stable state transition and ISC transition frequencies (Fig. 4e) are significantly different for emitters which have almost identical ZPLs. For ZPLs 660 nm, 657 nm, 630 nm, 637 nm we get for $R_0^{31}$ - 2.53±0.42 kHz,

0.59±0.10 kHz, 2.12±0.20 kHz, 1.26±0.34 kHz and for $R^{23}$- 2.95±1.88 kHz, 0.52±0.39 kHz, 4.49±1.46 kHz, 2.63±1.07 kHz, respectively. The emission frequency of emitters is described by the following equations:

$$C_{inf} = \frac{(R_0^{21}+R_0^{23})\phi_F}{2+\frac{R_0^{23}}{R_0^{31}}}, \quad C(P_{exc}) = C_{inf}\frac{\frac{P_{exc}}{P_s}}{1+\frac{P_{exc}}{P_s}}$$

Where $C_{inf}$ is the saturated emission frequency, $C(P_{exc})$ is the power dependent emission frequency, $\phi_F$ the quantum yield and $P_s$ is the saturation power. Our meta-stable and ISC rates are all in the kHz scale whereas the excited state rates are in the MHz scale, meaning $R_0^{21} \gg R_0^{31}, R^{23}$. Therefore, although the meta-stable and ISC rates are diverse, the dominant factors affecting emission efficiency are the excited state transition rates and the quantum yield. We can ascribe these diversities to different local environments in which the SQEs are embedded, as was observed for e.g. NV defect in nano-diamonds [31] or perhaps due to different charge states of h-BN, which were recently demonstrated for defects in h-BN.[32] The diversity of meta-stable and ISC transition rates, in account with the roughly homogenous environment in the monolayer, are another strong indicator to the possible presence of charge states as will be discussed further.

For emitter (639 nm ZPL) the excited state relaxation rate is $R_0^{21}$=620±140 MHz (Fig. 4g). Fig. 4h shows the different $\lambda_2$ frequencies for two CW excitations: 532 nm (green) and 594 nm (orange). The two curves converge upon zero power indicating that the transition frequency is not affected in the meta-stable state. We get $R_0^{31}(594\text{ nm}) = 3.5 \pm 0.3$ MHz and $R_0^{31}(532\text{ nm}) = 3.2 \pm 0.6$ MHz which are well in the range of each other. However, for the ISC we get different rates: $R^{23}(594\text{ nm}) = 27.5 \pm 1.9$ MHz and $R^{23}(532\text{ nm}) = 13.4 \pm 4$ MHz. The ISC transition rate for 594 nm excitation is approximately a factor of 2 larger than for 532 nm. Therefore, 532 nm leads to an efficient de-shelving of the metastable state. This can be seen when comparing the emission counts using the two different excitations (Fig. 4f), revealing that emission starts to saturate as we leave the linear regime behavior. Therefore, we extrapolate the saturation curves using $C_{inf}$ and C, demonstrating 532 nm is more efficient (Fig. 4f).

To further exclude the source of the defects as external ad-atoms we also attempted to expose exfoliated flakes to different gas environments and different chemical environments. An analysis can be seen in the SI, showing that emitters exhibit the same spectral features after treatment. Previous research using a scanning tunneling microscope has also shown that charged h-BN defects are intrinsic and not caused by ad-atoms.[32,33] We therefore sought out to ascertain if the SQEs in h-BN are spatially correlated with the h-BN hexagonal structure. Exfoliated h-BN exhibits more uniform crystallinity. To this end, we analyze an example of two emitters on the same confocal spot and between pairs of emitters on the same flake in close proximity by rotating the polarization of the excitation laser. This analysis is displayed in the SI and shows that emitter orientation is related to the h-BN hexagonal structure. We note that the large number of grain-boundaries in CVD h-BN excluded the option to check the orientation of emitters relative to each other in CVD h-BN in the same fashion. The spectral features we have observed, where the peak energy difference fits the splitting between longitudinal (LO) and transverse optical phonons (TO), is another fingerprint for a defect embedded inside the lattice, as seen also for SQEs in the UV range.[16]

We now reflect upon the possible effects of sample preparation methods presented. Carbon impurities were shown to account for UV luminescence [15] and recently for SQEs in the UV range, substituting the N vacancy site.[16] The chemical treatments and Ar/H$_2$ environments (see SI) the flakes were exposed to, reduce the probability that organic contaminants are responsible for the emitters seen, unless a carbon impurity was embedded inside and not exposed to the external areas of treatment. Since samples were annealed at 850 °C, oxygen should desorb from the flake (above 300 °C [34]) as well as gas atoms possibly trapped at interfaces. The suspended h-BN flakes also discard trapped atoms. We thus conclude that it is unlikely that organic additives or trapped gas atoms are responsible for the emitters found. Nevertheless, the chemical treatments done also have a strong oxidizing effect on the flakes [22] and can incorporate oxygen inside the lattice which cannot be discarded. We find that He irradiation generates the highest density of emitters (Table 1). Approximating the He atom inelastic collision to an electron inelastic collision we can compare to TEM research.[35,36] Temperature variation during irradiation creates different defects shapes [36] demonstrating that defect formation is dependent on the environment around the defect, emphasizing the importance of using gas during the annealing process compared to vacuum and could explain the difference in emitter stability. The vacancy sizes created cannot only be point-like (atomic) but also nano-sized as seen in the TEM images of Refs. 35,36. Annealing at 850 °C may cause vacancies to diffuse inside the flake [37] similar to diffusion known for the NV defect annealing in diamond,[38] leaving defects not exposed to the external environment. However, for the mono-layer, the defect is directly exposed to the environment.

Previously, the ZPL of 623 nm was attributed to the anti-site complex $(N_B V_N)$.[21] Our data shows the ZPL emission line to be between 1.78 eV to 2.18 eV (569 nm to 697 nm), suggesting that other types of impurities could also be responsible for the observed emission lines. The structural features we have observed indeed suggest a link between structure and defects. We discuss possible links comparing TEM data from previous research and density functional calculations. We note that clear identification of defect structure requires ultra-clean conditions due to the tendency of adsorbents (carbon and oxygen) to substitute defective sites,[28,39] possibly also accountable for emission.[40]

HRTEM has shown interlayer bonding on flake edges (nano-arches),[41] which can also be induced by electron irradiation of h-BN.[42] This can explain our observations of emitters located at perimeters of a flake on top of another flake (Fig. 1f/1g) or at boundaries (Fig. 1e), for untreated flakes and chemically treated flakes, as shown for the SQE in Fig. 2a/2b. It can also explain our irradiation results.

The B vacancy (visualized in TEM [43,44,45,46]) has a high average density of states in the energy range of 1.78 to 2.18 eV,[47] making it a possible candidate. However, the dynamics behind vacancies in 2D materials can be more complex than in 3D solids: B vacancies exhibit interlayer bonding behavior seen in bi-layer h-BN which do not occur for the N vacancy [45]: N bonds occur in the layer with the defect and one or two B atoms from the intact layer, reducing the symmetry of the vacancy from three- to two-fold,[44] consistent with the symmetry of the $N_B V_N$ complex. [21] This type of reconstruction is not observed for mono-vacancy defects inside multi-layered h-BN which retain their three-fold symmetry.[48] This hints at a connection between the low dimensionality of h-BN to quantum features [49] not seen *inside* bulk h-BN but occurring at the boundaries of the material, reminiscent to our observations.

Bonds not native to monocrystalline h-BN may play an important role for visible spectrum emission. Due to alternating bonds in h-BN (B-N), a non-native bond would be of the type of B-B or N-N, known as a homo-elemental bond which belongs in literature to Stone-Wales defects. These comprise a *family* of defects which also the $N_B V_N$ complex [21] is included in due to N-N-N bonds. A Stone-Wales defect can create square-octagon/pentagon–heptagon pairs in h-BN, changing bond lengths and having charge state configurations. Monolayer h-BN Stone-Wales defects have been visualized in TEM,[50,51] forming due to grain boundaries between domains, thus decreasing the band-gap and opening new energy states.[50] Calculated energy levels suggest emission in the visible spectral range [52] and with a less dominant phonon DOS of ~ 41 THz (~170 meV), fitting our TO and LO observations.[53] We can now link the angular multiplicity we have seen on perimeters where emitters are clustered (Fig. 1h/1g) as another hint to the presence of Stone-Wales defects or a grain-boundary effect.[52,53] Grain-boundaries in h-BN have been seen to be up to 90 μm long,[54] well in range of our perimeter lengths. Although $N_2$ ion irradiation has been shown to generate primarily N vacancy defects, they can also create interstitials which do not diffuse out of the lattice when annealing at 850 °C. These can create homo-elemental bonds not native to the lattice,[24] and hence also the similarity to monolayer h-BN. Similarly, local changes in h-BN layer stacking order has been shown via cathodoluminescence to account for a 1 eV range of emission of excitons (originating at ~200 nm wavelength).[13]

The quantum nature of emission and the high transition rates (> 500 MHz) measured for the excited states can arise from the reduced dimensionality of h-BN and the small bond lengths ( 1.4 Å [3]) which leads to large overlap between orbitals.[49] The variety of meta-stable and ISC transition frequencies found in CVD monolayer h-BN can have numerous causes. An intermittent charge state change during the acquisition of photon arrival times in our auto-correlation measurements not distinguishable on the photon trace, similar to the $NV^0$ and $NV^-$ charge states in diamond [55] would affect these transitions. Spin-orbit coupling [56] can also influence these transitions. Oxygen binding to nearby N vacancy sites [14,28] in proximity to the emitter can affect transition rates [57] or cause intermittency via charge tunneling.[58] It is worth noting that two-level and three-level systems have also been seen for the zinc vacancy in ZnO.[59]

To summarize, we investigated the correlation between different fabrication methods for defects in h-BN and the emitter density, photostability and structural features of h-BN. The creation of defects using chemical exfoliation further demonstrates the role of the exterior surface for QEs in h-BN. Our results demonstrate that the thinner and more edgy the flake is, the higher single emitter density becomes. We showed that slow rate chemical etching is a more facile method compared to ion irradiation for single emitter creation in multilayered exfoliated h-BN, since it can be more easily fine-tuned to a slow rate.[22] Using an insulating mask for ion irradiation might be beneficial to avoid ensemble creation by reducing collision damage of the ions. However, this would necessitate more steps in the process. Previous research has shown that by electron irradiation single emitters can be generated ,[19] the smaller collision profile of an electron can therefore be more beneficial as compared to an ion collision profile for emitter creation. Nevertheless, for inducing vacancy defects in monolayer h-BN, high energy ion irradiation can be a useful method, and using atoms with higher mass than nitrogen might be useful to tailor the type of vacancy created.[60]

We showed that similar emitter spectral features are present from bulk h-BN down to the monolayer. We also measured emitters' ZPL spectral FWHM comparable to that of bulk h-BN [21] and we found emitters which are photo-stable under orange laser excitation.

Our results show color diversity of h-BN consistent with previous reports [19,20], with ZPLs in the red visible spectrum, thereby enabling excitation with less energetic lasers, as was done in the h-BN CVD monolayer case. Using pyrolitic h-BN species might increase chemical inertness,[15,61,62] reducing free oxygen concentration and impurities in h-BN, thus eventually reducing spectral diffusion and photo-bleaching.

Further understanding of the origin of SQEs in h-BN can lead to more advanced fabrications methods in the CVD growth phase or after the growth phase, using ion irradiation and ion implantation, which has been demonstrated for Si [63], Be [64] and Ar [65] - embedded in the lattice. Band-gap tuning can be realized using monolayer h-BN as a platform with other monolayer 2D materials, tailoring the emitter's energy levels inside the band-gap. Clearly, a wide range of parameter adjustment can lead to diverse SQE defects in this wide band-gap material.

The work presented here paves the way for deeper understanding of the origin of QEs in h-BN. It proposes two fabrication methods to create emitters, demonstrating the flexibility of the material and emphasizing the role structure has on defect dynamics in 2D materials. These results highlight the diversity in the new arising field of 2D quantum emitters.

**ASSOCIATED CONTENT**

Supporting information contains (S1) more examples of emitter relation to structural features, (S2) an SEM image of CVD monolayer h-BN, (S3) details on chemical etching methods, (S4) details on ion irradiation on PL spectral features found for emitters using this method, (S5) excitation polarization measurements for emitter pairs, (S6) PL spectrum after different chemical and annealing environments, (S7) in-house grown CVD h-BN, (S8) detailed $g^2(t)$ measurement for an emitter.

**Corresponding Author**

*Inquiries should be sent to the following e-mail address: a.finkler@physik.uni-stuttgart.de

**Author Contributions**

All authors contributed to writing the paper. The authors acknowledge support from the Max Planck Society as well as the EU via the project DIADEMS and the DFG. AF acknowledges financial support from the Alexander von Humboldt Foundation. JHS acknowledges financial support from the EU graphene flagship.

**Acknowledgments**

The authors thank Sang-Yun Lee, Durga Dasari, Denis Antonov, Marcus Doherty for valuable discussions, Matthias Widmann for technical support, Andrea Zappe for chemical treatment assistance and Johannes Greiner for mathematical computational assistance.

**ABBREVIATIONS**

h-BN, hexagonal boron nitride; ZPL, zero phonon line; SQE, single quantum emitters; PL, photo-luminescence; CVD, chemical vapor deposition; APD, avalanche photo diode; CW, continuous wave; FWHM, full width half maximum; TO/LO phonons, transverse/longitudinal optical phonons; ISC, inter system crossing; HR/TEM, high resolution/ transmission electron microscope; DOS, density of states.

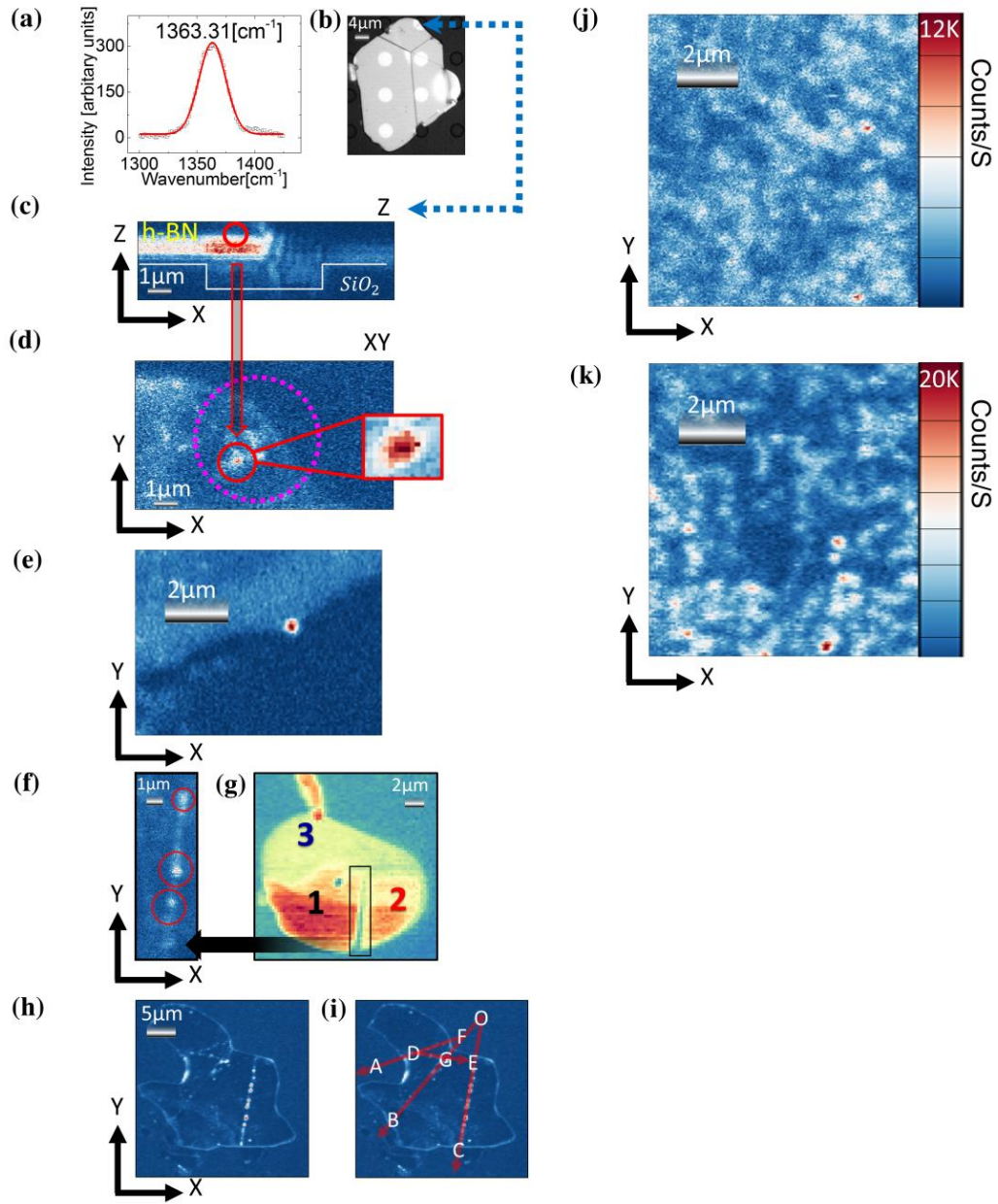

Figure 1:

(a) Raman spectra confirming h-BN of bulk scale on 180 μm thick $SiO_2$. –(b) A microscope image of a flake on $SiO_2$ with holes is shown. (c)(d)(e)(f)(g)(h)(i)(j)(k) Confocal scans of exfoliated flakes, color coding indicates counts per second. (c) Z axis, (d) XY axis of an area of the same flake as in (b) (indicated by the blue dashed arrow), suspended over a hole (marked with a purple circle). (c) is unfiltered and (d) is filtered. For unfiltered scans, the laser power was substantially reduced (nW) to avoid burning the APDs and the wavelength filters were removed. A suspended emitter is seen (marked in red). (e) QEs appearing at the boundary (f)(g) and interface

(filtered/unfiltered, respectively) (g) of a flake on top of another, different flake. Thicknesses are indicated (1 to 3). (h) An angular multiplicity of 30 degrees is found for emitters seemingly aligned along straight lines on an exfoliated flake. (j)(k) 532 nm CW confocal scan for a comparison between a chemically etched flake (j) and CVD h-BN (k)

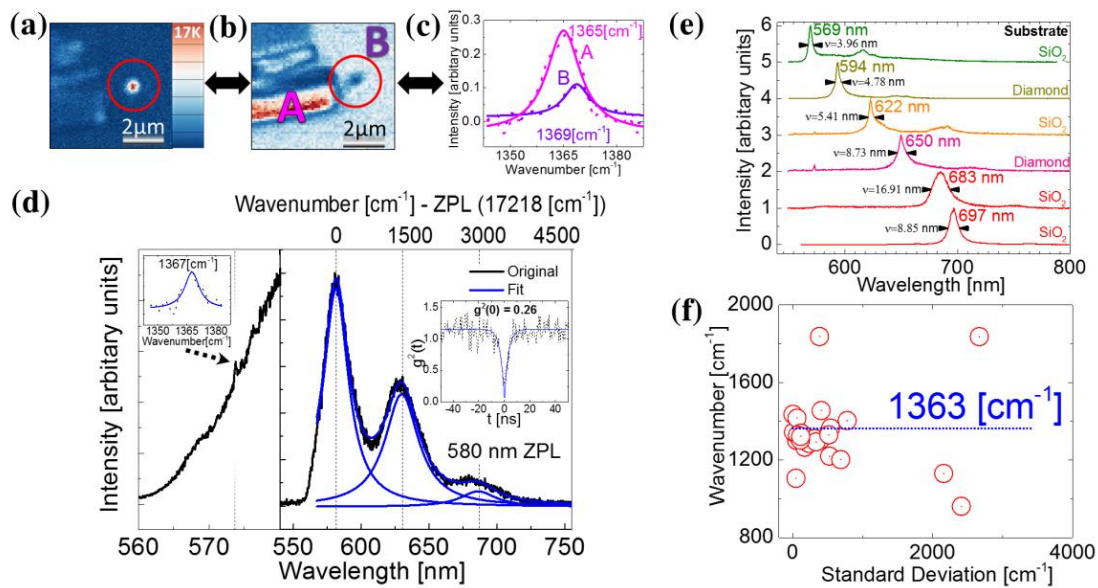

Figure 2:

(a) Filtered (b) unfiltered confocal scan of a SQE, colored scale bar indicates counts per second, (c) Raman spectroscopy was done on areas marked with A and B to verify different thicknesses. (d) Left – A high resolution PL spectrum of the SQE using 532 nm CW excitation, showing the typical Raman signal superimposed on the defect spectral features (indicated by the dashed black arrow). Right – Low resolution PL spectrum (black curve). The de-convoluted spectra are indicated in blue. Inset shows an autocorrelation measurement of the SQE. (e) PL spectrum using 532 nm CW excitation from h-BN on different substrates (indicated for each curve). ν denotes the ZPL full width half maximum in nm. All of these six representative spectral features were seen both on diamond and SiO$_2$. (f) Conversion of 23 emitter PL spectrum acquired from various sources from treated/untreated flakes to the difference in wavenumber between adjacent peaks as a function of standard deviation.

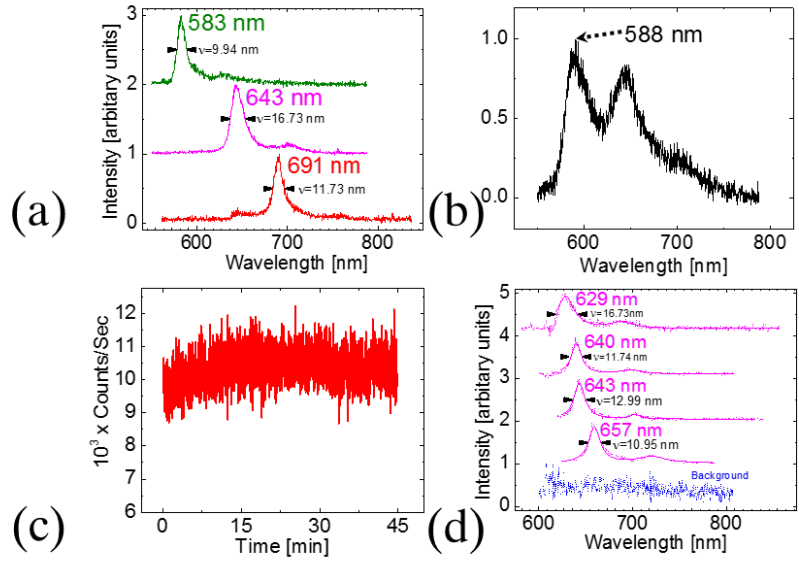

Figure 3:

(a)(b) CVD mono-layer PL spectra using 532 nm CW excitation (a) PL spectra for 3 emitters (b) PL spectra for a suspended emitter. (c)(d) 594 nm CW excitation using a 600LP filter (c). Emission count from an emitter with a 640 nm ZPL for 45 min (d) PL spectra for 4 emitters.

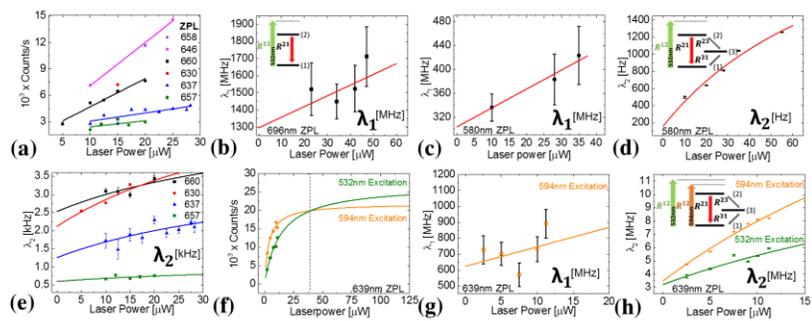

Figure 4:

(a) Photon counts for emitters in CVD mono-layer h-BN divided into two categories: two-level systems (658 nm and 646 nm) and three-level systems (660 nm, 630 nm, 637 nm and 657 nm) for 594 nm excitation. Power dependent transition rates for 696 nm ZPL (b), 580 nm ZPL (c)(d) and 639 nm ZPL (g)(h). (f) Photon counts for the 639 nm emitter, for 594 nm excitation and 532 nm excitation. Saturation curves were fitted as described in the text. The dashed line indicates crossing of the two curves (e) Meta-stable and ISC transition rates for 4 emitters with ZPLs 660 nm (black), 630 nm (red), 637 nm (blue) and 657 nm (green) in CVD mono-layer h-BN for 594 nm excitation.